\pdfoutput=1
\documentclass[prl,twocolumn,floatfix,superscriptaddress]{revtex4-1}
\usepackage{times,amsmath,amssymb,amstext,latexsym,float,graphicx,color}
\usepackage{bm}
\usepackage{hyperref}
\hypersetup{colorlinks=true, citecolor=blue, urlcolor=blue, linkcolor=blue}

\begin{document}
	
\title{Symmetry-controlled  singlet-triplet transition in a double-barrier quantum ring}

\author{Heidi Potts}
\email[]{heidi.potts@ftf.lth.se}
\affiliation{Division of Solid State Physics and NanoLund, Lund University, SE-221 00 Lund, Sweden}
\author{Josef Josefi}
\affiliation{Mathematical Physics and NanoLund, Lund University, SE-221 00 Lund, Sweden}
\author{I-Ju Chen}
\affiliation{Division of Solid State Physics and NanoLund, Lund University, SE-221 00 Lund, Sweden}
\author{Sebastian Lehmann}
\affiliation{Division of Solid State Physics and NanoLund, Lund University, SE-221 00 Lund, Sweden}
\author{Kimberly A. Dick}
\affiliation{Division of Solid State Physics and NanoLund, Lund University, SE-221 00 Lund, Sweden}
\affiliation{Centre for Analysis and Synthesis, Lund University, SE-221 00 Lund, Sweden}
\author{Martin Leijnse}
\affiliation{Division of Solid State Physics and NanoLund, Lund University, SE-221 00 Lund, Sweden}
\author{Stephanie M. Reimann}
\author{Jakob Bengtsson}
\affiliation{Mathematical Physics and NanoLund, Lund University, SE-221 00 Lund, Sweden}
\author{Claes Thelander}
\email[]{claes.thelander@ftf.lth.se}

\affiliation{Division of Solid State Physics and NanoLund, Lund University, SE-221 00 Lund, Sweden}

\date{\today}

\begin{abstract}
We engineer a system of two strongly confined quantum dots to gain reproducible electrostatic control of the spin at zero magnetic field. Coupling the dots in a tight ring-shaped potential with two tunnel barriers, we demonstrate that an electric field can switch the electron ground state between a  singlet and a triplet configuration. Comparing our experimental co-tunneling spectroscopy data to a full many-body treatment of interacting electrons in a double-barrier quantum ring, we find excellent agreement in the evolution of many-body states with electric and magnetic fields. The calculations show that the singlet-triplet energy crossover, not found in conventionally coupled quantum dots, is made possible by the ring-shaped geometry of the confining potential. 
\end{abstract}

\maketitle

The ability to engineer electron spin states is key for a wide range of applications, from sensing to quantum computation~\cite{Loss1998,Awschalom2013,Kloeffel2013}. It is also important for developing a microscopic understanding what role magnetic impurities play in electron transport ~\cite{Bulla2008}. Configurations involving several spins, such as spin-singlet and triplet  states, have provided new ways to operate spin qubits~\cite{Petta2005,Maune2012,Reed2016}, and to gain new insight to, for example, quantum criticality~\cite{Roch2008}, and Cooper pair transport in Josephson junctions~\cite{Estrada2018,Bouman2020}. 

The possibility to distinguish individual spin states depends on their energy splitting. While it can be controlled with magnetic fields (Zeeman effect), electrostatic tuning is of interest for fast and selective operations. However, achieving full electrostatic control over the electronic structure and spin of a nanoscale system, such as a quantum dot (QD), is a challenging endeavor~\cite{Reimann2002,Hanson2007}. The typically large orbital separation in QDs often leads to a stable spin-singlet ground state (GS). In the case of orbital degeneracy, the Coulomb repulsion between the electrons may instead favor a spin-triplet GS due to its antisymmetric spatial wave function (Hund’s rule). A triplet GS has been observed in QDs with either symmetric or very weak confinement such as in GaAs 2DEGs~\cite{Tarucha1996,Schmid2000,Kogan2003,Martins2017,Malinowski2018}, a C$_{60}$ molecule~\cite{Roch2008}, carbon nanotubes~\cite{Liang2002,Quay2007} and quasi-1D ring potentials~\cite{Warburton2000,Fuhrer2003}. Reports on electrostatically induced singlet-triplet GS transitions are however only a handful. They hitherto rely on gate-induced shifts of orbital state energies or tunneling probabilities~\cite{Kyriakidis2002,Kogan2003,Fuhrer2003,Quay2007,Roch2008,Malinowski2018}, which are difficult to engineer in a single QD.

Double QDs (DQDs) provide a more direct way to control the confining potential and the spin states of a few interacting electrons. In the case of even electron numbers, the singlet-triplet energy splitting  $\Delta _{T-S} = E_T - E_S$, given by the energy difference of the triplet ($T$) and singlet ($S$) states, can be tuned by the inter-dot tunnel coupling~\cite{Hatano2008,Nilsson2018}. However, delocalization of the electrons favours a spin-singlet GS in a conventional DQD, and $\Delta _{T-S}$ is therefore positive~\cite{Burkard1999}.

\begin{figure}[b]
	\centering
	\includegraphics[width = \columnwidth]{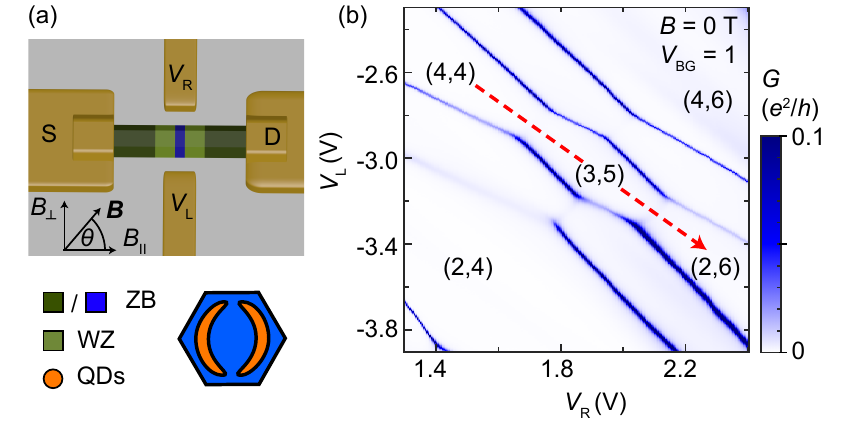}
	\caption{{\it ({Color online})} (a) Illustration of the device and the formation of two quantum dots near the nanowire surface. (b) Conductance $G$ as a function of side-gate voltages $(V_{\mathrm{L}},V_{\mathrm {R}})$ for Device-I ($V_{\mathrm{sd}}=0.3$\,mV). The electron number on each quantum dot is labeled by the charge configuration tuple $(N_{\mathrm{left}}, N_{\mathrm{right}})$.   The red (dashed) arrow indicates detuning from the $(4,4)$ the $(2,6)$ charge configuration of the DQD.}
	\label{main:fig1}
\end{figure}

In this Letter, we show that a DQD shaped like a quantum ring can have a spin triplet ground state for electron numbers $N=4j$, where $j$ is an integer, and that the ground state spin can be manipulated by electric fields. We investigate a system based on InAs nanowires where epitaxial barriers confine electrons to a thin disc. Using co-tunneling spectroscopy, we show that a symmetry-breaking electric field induces a triplet-singlet ground state transition, which is reproduced among many samples and electron configurations. We confirm the experimental results with a full many-body calculation of Coulomb-interacting electrons in a double-barrier quantum ring.

The few-electron spin states reside near the surface of a disc-shaped zinc-blende (ZB) section of an InAs nanowire, strongly confined between two wurtzite (WZ) sections acting as tunnel barriers, as illustrated in Fig.~\ref{main:fig1}(a). Tuning a pair of side-gates ($V_\mathrm{L}, V_\mathrm{R}$) and a global back-gate $V_{\mathrm{BG}}$ allows to change the confinement potential such that two QDs are created which are parallel-coupled to source and drain. Thus, the number of confined electrons can be controlled down to the last electron~\cite{Nilsson2017,Nilsson2018,Thelander2018,Potts2020}. A magnetic field 
with magnitude $B$ is applied in the $(y,z)$-plane with angle $\theta $ to the $z$-axis. Hence, with the ZB disc in the $(x,y)$-plane, for $\theta =0 $ the field is aligned parallel ($B_{\mid\mid }$) to the long axis of the nanowire, and for $\theta = \pi/2$ it is perpendicular to it (${B_{\perp }}$).

The conductance $G$ as a function of side-gate voltages $V_\mathrm{L}$ and $V_\mathrm{R}$ for Device-I is shown in Fig.~\ref{main:fig1}(b). The parallel DQD behavior manifests itself in a characteristic honeycomb pattern~\cite{Wiel2002}, where we can label the number of electrons on the left and right QD by the  charge configuration tuple $(N_{\mathrm{left}}, N_{\mathrm{right}})$. We first investigate spin states involved in transport in the $(3,5)$ honeycomb. The gate voltages are optimized such that the two QDs of the DQD system are coupled at two locations, resulting in a quantum ring structure with two tunnel barriers, as demonstrated in Ref.~\cite{Potts2019}. 

\begin{figure}[t]
	\centering
	\includegraphics[width = \columnwidth]{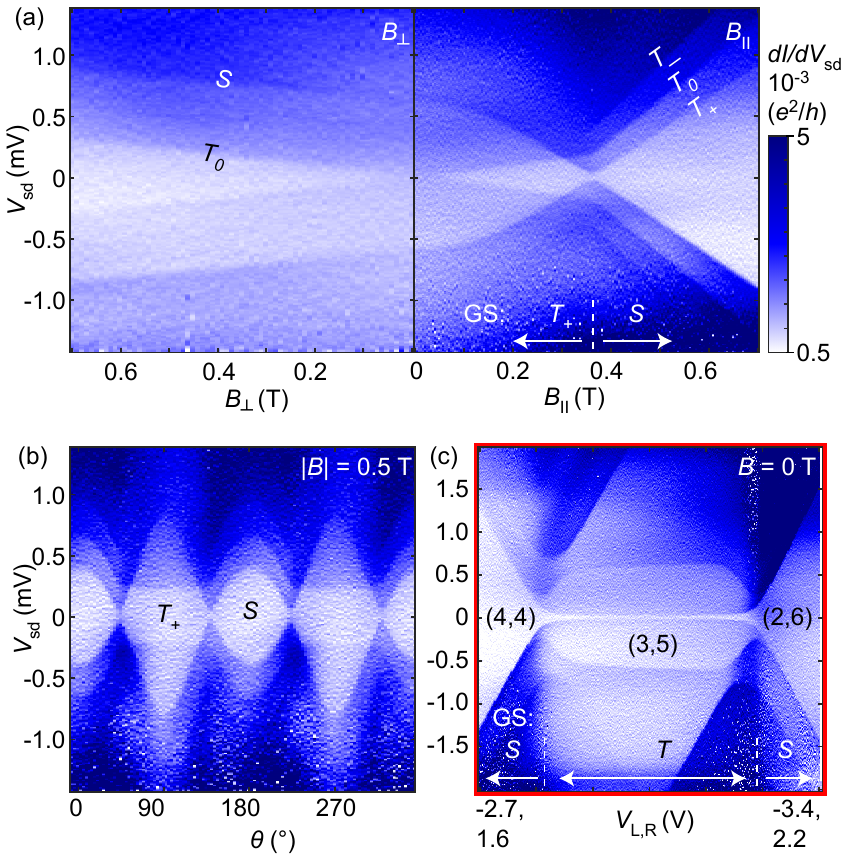}
	\caption{{\it ({Color online})}
		(a) ${\mathrm{d}}I/{\mathrm{d}}V_{\mathrm{sd}}$ versus $V_{\mathrm{sd}}$ in the centre of the $(3,5)$ honeycomb structure as a function 
		of ${B_{\perp }}$ and $ B_{\mid\mid }$. A singlet-triplet splitting $\Delta _{T-S} \approx -600~\mu \mathrm{eV}$ is extracted at $B=0$. The ground state and the excited states (visible as lines marking abrupt changes in ${\mathrm{d}}I/{\mathrm{d}}V_{\mathrm{sd}}$) are labelled based on the theoretical data in Fig.~\ref{main:fig2}. (b) ${\mathrm{d}}I/{\mathrm{d}}V_{\mathrm{sd}}$ versus $V_{\mathrm{sd}}$  as a function of ${\bf B}$-field direction for $  B  = 0.5 ~\mathrm{T}$. (d) ${\mathrm{d}}I/{\mathrm{d}}V_{\mathrm{sd}}$ versus $V_{\mathrm{sd}}$  as a function of detuning along the red dashed gate vector (shown in Fig.~\ref{main:fig1}(b)) for $B = 0~\mathrm{T}$.
	}
	\label{main:fig2}
\end{figure}

\begin{figure}[]
\centering
\includegraphics[width = \columnwidth]{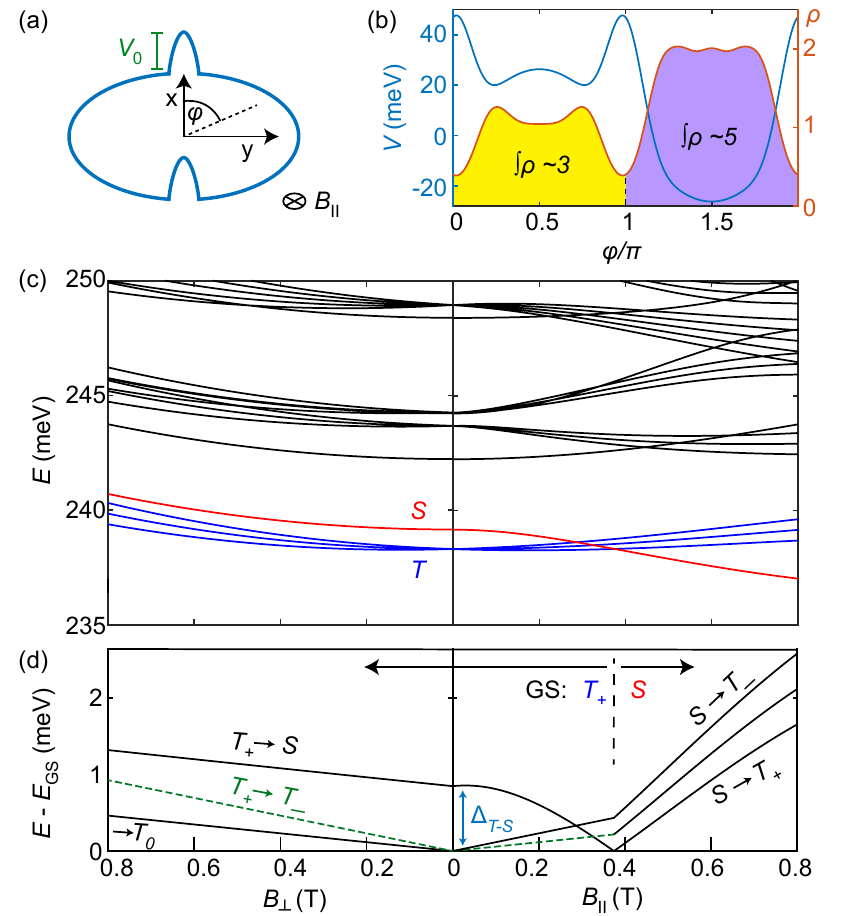}
\caption{{\it (Color online)} (a) Sketch of the one-dimensional ring potential at zero detuning. (b) Optimized ring potential and single-particle density distribution $\rho$ as a function of azimuthal coordinate $\varphi $. (c) Eigenstate energies as a function of $B_{\perp }$ and $B_{\mid\mid }$. (d) Energy difference between the ground and excited states. Note that the transition from $T_+$ to $T_-$ (green dashed line) requires two spin flips which is not possible in a single co-tunneling process.
}
\label{main:fig3}
\end{figure}

Figure~\ref{main:fig2}(a) shows differential conductance, ${\mathrm d} I/\mathrm{d} V_{\mathrm{sd}}$, as a function of source-drain voltage, $V_{\mathrm{sd}}$, and magnetic fields $B_{\mid\mid }$ and ${B_{\perp }}$. Due to the large charging energy and orbital level spacings,  all transport features are related to (inelastic) co-tunneling, involving excited states of the same charge configuration (3,5). At $B=0$ we observe conductivity due to co-tunneling at $V_{\mathrm{sd}}=0$, as well as the onset of an additional co-tunneling pathway due to an excited state  at $V_{\mathrm{sd}} \approx 0.6~{\mathrm{mV}}$. Supported by the calculations shown in Fig.~\ref{main:fig3}(c) and (d), we identify the GS at $B=0$ as a spin triplet ($T$),  and the first excited state  as a spin singlet ($S$). Independent of the ${\bf B}$-field direction, the triplet GS spin-splits with increasing magnetic field. However, the energy gap to the singlet excited state increases with increasing $B_{\perp }$, while it decreases with $B_{\mid\mid}$. Consequently, $S$ becomes the GS at $B_{\mid\mid} \approx 360~\mathrm{mT}$, and three triplet states $(T_+, T_0, T_-)$ are observed as excited states. An in-plane rotation of a fixed-magnitude magnetic field can therefore result in a periodic change of the $N=8$ electron GS as shown for $B = 0.5~\mathrm{T}$ in Fig.~\ref{main:fig2}(b).

Figure~\ref{main:fig2}(c) shows the bias-dependent differential conductance when detuning the DQD from the $(4,4)$ to the $(2,6)$ charge configuration, as indicated by  the red dashed arrow shown in Fig.~\ref{main:fig1}(b). We find that $\Delta _{T-S}$ is independent of detuning in the $(3,5)$ regime, confirming that the singlet and triplet states are composed of the same charge configuration. 

We now turn to a theoretical explanation of the singlet and triplet states and their magnetic-field dependence. Except for magnetic fields close to zero, the ring-shaped DQD states behave qualitatively similar to those of a simple quantum ring without tunnel barriers. Assuming non-interacting electrons, they would be distributed between states of orbital angular momentum $l_z$ for the case of a ring. For $N=8$, the lowest energy triplet state has a closed-shell core with six electrons (two electrons with each $l_z=-\hbar, 0,$ and $\hbar$), plus one electron with $l_z = - 2\hbar$ and one with $2\hbar$. Hence, the triplet state has total angular momentum $L_z=0$. The lowest energy singlet state has the same closed-shell core as the triplet one but instead two electrons of opposite spins with $l_z=-2\hbar$ and thus in total $L_z = -4\hbar$. By increasing $B_\parallel$, the singlet therefore decreases in energy relative to the triplet. Because of the azimuthal symmetry and the orbital degeneracies, a similar behavior is expected for $N=4j$.
Such a simple model provides an intuitive understanding of the observed magnetic field dependence of the singlet and triplet states. It has previously been employed to describe even-electron states in carbon nanotubes ~\cite{Jespersen2011}.

However, this simple picture gives no indication to the magnitude of $\Delta _{T-S}$. It also does not capture the effects from the two barriers within the ring, such as the reduction in $L_z$ at small $B$. Experimentally we also find that a spin triplet GS requires odd electron numbers in each of the QDs  in addition to $N=4j$. In order to explain these findings, and the electrostatic tuning of the ground state spin, we performed a full many-body study of Coulomb interacting electrons in a double-barrier quantum ring. The $N = 8$ electron system in the $(3,5)$ regime is modeled using a structured (one-dimensional) ring confinement with radius $R = 23~\mathrm{nm}$ in the $(x,y)$-plane. In particular, we consider two potential barriers opposite to one another, effectively creating a double half-ring geometry, see Fig.~\ref{main:fig3}(a). An in-plane electric field is controlled by the side-gate potential $V_{\mathrm g}$. The resulting effective confinement potential is approximated as 
$V=V_0 |\cos (\varphi )| ^q + V_{\mathrm{g}} \cos (\varphi - \varphi _{\mathrm{g}})$.  
Here, $\varphi $ is the azimuthal angular coordinate, $V_0$ is the barrier height,  the exponent $q$ 
determines the barrier width, and $\varphi _{\mathrm{g}}$ specifies the direction of the electric field relative to the position of the barriers. In a homogeneous magnetic field, the kinetic momentum of the electrons equals 
${\bf p}+e{\bf A}$, where ${\bf p}$ is the canonical momentum, and $e$ is the elementary charge. 
Here, with the usual Coulomb gauge, we write ${\bf A}=({\bf B}\times {\bf r})/2$ where ${\bf r}$ is the position vector. 
For the one-dimensional confinement, we regularize the electron-electron Coulomb interactions and retrieve the eigenenergies and -states to the system from the diagonalization of the full many-body Hamiltonian, 
using the configuration interaction (CI) method~\cite{Reimann2002} (Supplemental Material). We assume an effective electron mass $m^*=0.023m_e$ (where $m_e$ is the bare electron mass), a relative permittivity $\epsilon_r  = 15.15$ and an effective electron spin $g$-factor of $g^*_{\mathrm {spin}} = 10$. The parameters of the confinement $V$ are adjusted to match the experimental observations. In particular, for $B=0$ the many-body ground state has on average three electrons in one half-ring and five in the other, see Fig.~\ref{main:fig3}(b). Note that here we choose $\varphi _g=\pi /2$, such that  the confinement  is symmetric about the $y$-axis.

The evolution of the eigenstates calculated as a function of $B_{\perp }$ and $B_{\mid\mid }$ is shown in Fig.~\ref{main:fig3}(c). In line with their magnetic-field behavior, we label the states as spin-singlet (S) and spin-triplet (T), as introduced previously. At $B=0$, the system has a degenerate spin-triplet GS, and obeys Hund’s rule. Exposing the DQD to a magnetic field, we note three separate effects, see Fig.~\ref{main:fig3}(c): (i) There is a lifting of the spin degeneracy of the triplet ground state, which occurs regardless of the orientation of the magnetic field.  (ii) The diamagnetic term gives an overall shift of energies. (iii) When the field is oriented with ${ B_{\mid\mid }}$, there is  an additional and complicated dependence of the energy levels on the magnetic field strength, where certain states show a larger change in their energy than others. In particular, the lowest-energy singlet  state  decreases its energy with  $B_{\mid\mid }$ due to a significant orbital angular momentum, and becomes the GS for $B_{\mid\mid } > 375 {\mathrm{mT}}$. However, the system's orbital angular momentum $L_z$ is not a conserved quantity, i.e. for a single state its average angular momentum $L_z$ changes with increasing  $B_{z}$ due to the barriers in the ring. At the same time, also the kinetic, potential and interaction energy contributions vary with $B_z$. The interplay between these contributions results in a more complex magnetic field dependence, such as the flat dispersion of the lowest-energy singlet state for small magnetic fields (indicating that $L_z = 0$), which stands in contrast to a simple ring confinement (Supplemental Material).

\begin{figure*}[]
	\centering
	\includegraphics[width = \textwidth]{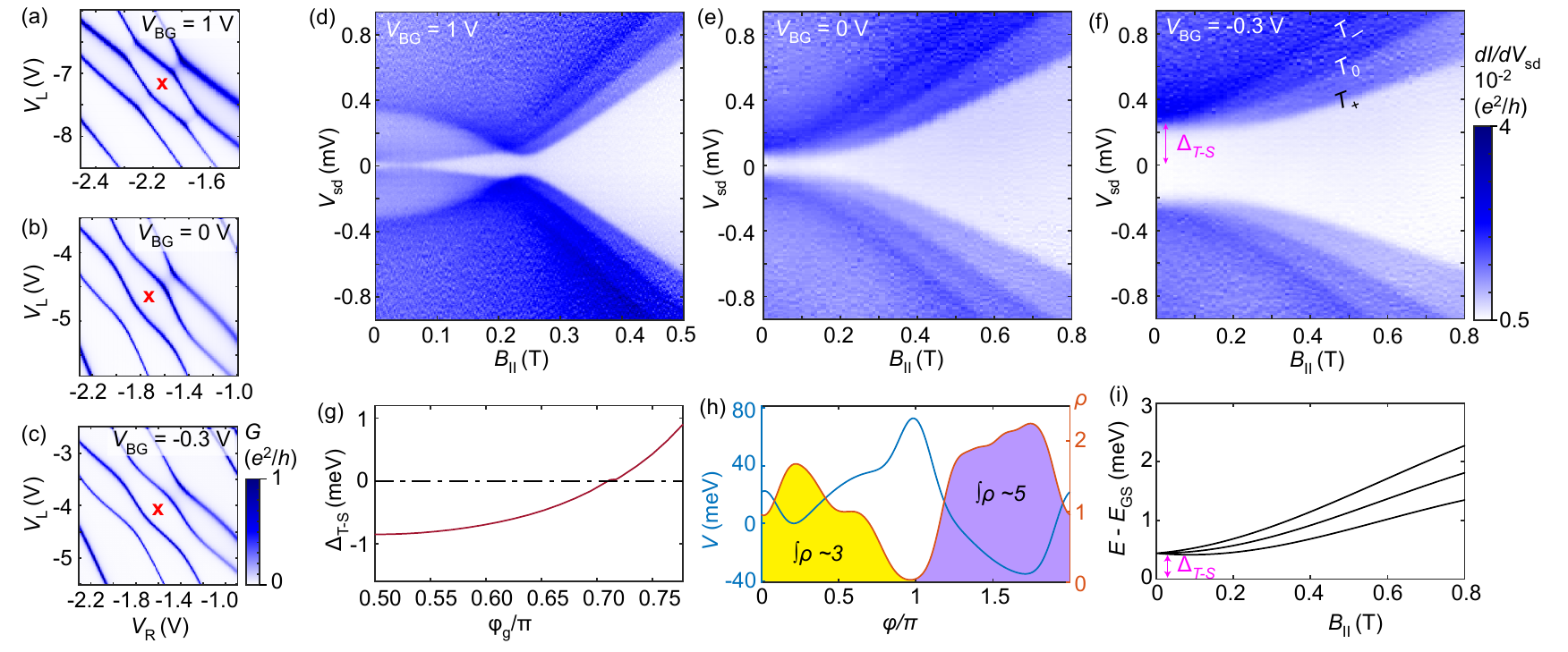}
	\caption{{\it (Color online)} (a)-(c) Conductance as a function of side-gate voltages for Device-II at 
		$V_{\mathrm{BG}} = 1\,{\mathrm V}, 0\,{\mathrm V},$ and $-0.3\,{\mathrm V}$ ($V_{\mathrm{sd}}=0.24$\,mV). (d)-(f) ${\mathrm{d}}I/{\mathrm{d}}V_{\mathrm{sd}}$ versus $V_{\mathrm{sd}}$
		in the centre of the honeycomb as a function of $B_{\mid\mid }$ for the same 
		$V_{\mathrm{BG}}$ values as in (a)-(c). $\Delta _{T-S}$ increases from $-340~\mu \mathrm{eV}$  to 
		$70~\mu \mathrm{eV}$ to $250~\mu \mathrm{eV}$. (g) Calculation of  $\Delta _{T-S}$ as a function of the angle $\varphi _g$ of the electric field relative to the barrier positions. (h) Ring potential and single particle density distribution as a function of the azimuthal coordinate $\varphi $ for $\varphi _g=0.75\pi $. (i) Energy difference between ground and excited states as a function of  $B_{\mid\mid }$.
	}
	\label{main:fig4}
\end{figure*}

To facilitate the comparison with the experimental measurements in Fig.~\ref{main:fig2}(a), we also plot the excitation energies relative to the GS energy in Fig.~\ref{main:fig3}(d). We note that the model does not include spin-orbit-interaction, which explains the exact singlet-triplet crossing in contrast to the avoided crossing observed experimentally.

In the following, we show how the GS can be tuned from triplet  to singlet by an electric field. In Fig.~\ref{main:fig4}(a-c) we plot the conductance of Device-II as a function of $V_\mathrm{L}$ and $V_\mathrm{R}$, for three different back-gate voltages. The side-gate voltages were here adjusted to keep the same charge configuration (3,5). Transport as a function of $V_{\mathrm{sd}}$ and $B_{\mid\mid }$  in the center of the honeycomb (marked with red crosses) is shown in Figs.~\ref{main:fig4}(d-f). Similarly to Device-I, a triplet GS can be observed at zero magnetic field  for 
$V_{\mathrm{BG}} = 1~\mathrm{V}$ as shown in Fig.~\ref{main:fig4}(d). Decreasing $V_{\mathrm{BG}}$ results in a round-off of the corners of the honeycomb structure, as shown in Figs.~\ref{main:fig4}(b,c) for $V_{\mathrm{BG}} = 0~\mathrm{V}$ and  $V_{\mathrm{BG}} = -0.3~\mathrm{V}$, and the evolution of states with $B_{\mid\mid }$ exhibits a clearly modified behavior (Figs.~\ref{main:fig4}(e,f)). The GS at $B=0$  in Fig.~\ref{main:fig4}(e) is a singlet, and the first excited state is a triplet which splits into three states for $B > 0$. Further decreasing the back-gate voltage increases the singlet-triplet splitting $\Delta _{T-S}$ in Fig.~\ref{main:fig4}(f). 

To confirm the observed transition from a triplet to a singlet GS, we accordingly modify the ring potential $V$ defined above, and again address the problem from the theoretical perspective. 
In particular, in Fig.~\ref{main:fig4}(g) we show that $\Delta _{T-S}$ can be controlled by $\varphi _g$, i.e. by changing the direction of the electric field relative to the position of the barriers. The amplitude of $V_{g}$  is also adjusted to ensure that three electrons always remain in one half-ring and five in the other, i.e. to mimic the experimental configuration.  Clearly, for $\varphi _g =\pi /2$, where the confining potential is symmetric along the $y$-axis, a spin-triplet many-body ground state is obtained (see also the discussion of Fig.~\ref{main:fig3}). Increasing $\varphi _g$ effectively enhances one of the barriers (and reduces the other one) and results in a less negative $\Delta _{T-S}$. Above the angle $\varphi _g =0.71 \pi $,  a spin-singlet many-body ground state appears. The confinement potential and the evolution of the states as a function of $B_{\mid\mid }$ for $\varphi _g =0.75 \pi $ are shown in Figs.~\ref{main:fig4}(h,i). We note that the lowest energy singlet has $L_z < 0$ at large B-fields despite the asymmetry in barrier transmission. This is very different from the case of a DQD coupled only in one location for which $L_z = 0$ for all states. A good qualitative agreement is found between the calculation in Fig.~\ref{main:fig4}(i) and the experimental observation in Figs.~\ref{main:fig4}(e,f).  We  emphasize that the favoring of the spin-singlet state can also be qualitatively understood from the natural orbitals of the system. In particular, at $\varphi _g = \pi /2$ there are three orbitals with occupation numbers $\langle n\rangle \sim 2$ (i.e. each orbital occupied by two electrons of different spin) and two with $\langle n\rangle \sim 1$. Here, the singly-occupied orbitals are each largely localized within a single half-ring. An increase in $\varphi _g$, however, leads to delocalized orbitals. The possibility for two electrons of different spin to instead occupy one of these spatial orbitals reduces the energy of the singlet state and here drives the  transition from a triplet to a singlet. (See also the Supplemental Material).

In summary, we have shown how to realize a strongly confined quantum system for which the electron spin configuration can be electrostatically controlled between singlet and triplet. The results are reproduced in several devices, fulfilling $N=4j$ and with an unpaired spin in each QD. The controllability is a consequence of combining the properties of a ring and a DQD. For a pure ring  geometry and for the considered electron numbers, a triplet ground state originates from Hund's rule at zero magnetic field. For a standard DQD, however, a singlet is preferred as a consequence of the lack of azimuthal symmetry.  A full many-body treatment of the interacting electron system confirmed a GS transition when increasing the size of one of the two tunneling barriers and thereby  transforming the ring-like geometry to a DQD-like one. The ability to tune the spin configuration of the ground state at zero magnetic field electrically makes such quantum ring structures an unique platform for spin-based applications.

\bigskip
\begin{acknowledgments}
{\it Acknowledgements.}  We thank A. Wacker for discussions. This work was supported by the Knut and Alice Wallenberg Foundation, the Swedish Research Council and NanoLund. H.P. thankfully acknowledges funding from the Swiss National Science Foundation (SNSF) via EarlyPostDoc Mobility Grant No. P2ELP2\_178221. \end{acknowledgments}

\bibliography{ST}
\end{document}